# Nyaya-Vaisheshika: The Indian Tradition of Physics

Roopa Hulikal Narayan

## 1 Introduction

This paper is the first in a series on the Indian tradition of physics that while summarizing the earlier review by Kak [1], [2] will set the stage for a more comprehensive analysis to follow in later papers. In ancient India, the schools of Nyaya and Vaisheshika focused on logic and atomic approach to matter. In this paper, the idea of atomicity and other physical ideas given in Vaisheshika are reviewed in light of the central role the observer plays in Indian thought. We provide introduction to ideas that are described in greater detail in Potter's text [10], where the focus is not on physical ideas but rather on philosophy.

The Rigveda, the oldest of the Vedic texts of India, generally assigned to the early second millennium BC or earlier, is seen within the Indian tradition as the source of its approach to reality. The Vedic sages recognized a binding unity among all that constitutes this universe. They made an attempt to reflect this pattern of interdependence among the entities of the universe including the very structure of universe itself. This may be seen in the structure and symbolic purpose of Vedic altars, approach to language, and so on [3],[4],[5]. The observer or the experiencing subject was given a privileged state in physical thought [6-10].

By the end of nineteenth century, the place of the observer also became a part of the mainstream discourse of academic physics and psychology in the consideration of the dichotomous issues of order and disorder. The second law of thermodynamics stood for the principle of increasing disorder in physics, whereas in biology, the theory of evolution is a principle of ever-increasing order and organization [6, page 50].

Parallels to this dichotomy occur in Indian thought: the self tends toward order whereas the atoms of the body it resides in and the mind it possesses tend toward disorder. But in the analysis of Indian writings we face the problem that its terminology is not always clear in commentaries. One of the tasks of the paper is to clarify the basic terminology used in the Indian physics tradition.

## 2 Nyaya-Vaisheshika

It is generally accepted [1],[10] that the origin of the Indian physical thought is in the Rig Veda where the order of nature is expressed as *Rta*. This *Rta* encompasses the laws of universe which are otherwise unexplainable unless an initial cause of universe is identified. The ability of consciousness to comprehend these laws is also in the purview of *Rta*. Kanada, one of the primary architects of Vaisheshika, states "I shall enumerate everything that has a character of being [1]".

Vasheshika analyzes material particles through two independent means of knowledge, which are recognized as Perception and Inference. The dependence on these two means alone is evident from the fact that the Vaisheshika accepts the authority of the Vedas based on inference alone [1]. The same principle of inference is added as a prelude to the inference of the self as the basis of cognitive states.

In India, objective science and the science of the self, go hand-in–hand. The Nyaya School as the discipline of logical inference complements Vaisheshika, and the two are often called as Nyaya-Vaisheshika.

## 3 Brief introductions to their Sciences

### 3.1 The Philosophers and their philosophy

All thought systems evolve with time, and the philosophy of Vaisheshika is no exception to this rule. The two primary philosophers of interest to us are Kanada and Prashastapada, because their contributions have been remarkably significant.

Kanada, one of the early philosophers of Vaisheshika, is known for his atomic view of the world. He uses the term '*Vishesha*' to mean particularity of an atom and also in the sense of '*Antya Vishesha*' meaning the 'final individuator' the ultimate individuality of each atom which individuates it from all else. This is a unique feature of this school and hence the term vishesha in its adjective form '*Vaisheshika*' is the name of this school.

Prashastapada, who came centuries after Kanada, describes the dissolution of earth, water, air and fire in terms of its atomic constituents that excludes space since its nature is taken to be non-atomic. The conjoining and disjoining of atoms is described as a natural property of atoms but Prashastapada includes a higher will (or order) as the guiding principle of universal dissolution which over-rides the natural karma (motion) of atoms [8, page 65].

The cosmological cycle of creation and dissolution at an atomic level and the breakdown of all natural properties of atoms at such a time of dissolution until the process of creation is re-begun when such natural properties hold good once again is a remarkable insight. Initially atoms were described to conjoin and disjoin resulting in creation and dissolution of new substances which is because of the karma (laws) [8, page 65].

### 3.2. Inference

#### 3.2. 1 Definition and Classification

Knowledge begins with cognition through two valid means of knowledge, namely Perception and Inference. Perception is when the sensory organs come in contact with some recognizable property of a substance like color of an object. Inference is the method of reasoning. It has a sub-classification of *Drshta* -what can be observed and, *Adrshta* – what cannot be observed.

*Adrshta* encompasses all that cannot be explained [7]. Although the inability to explain *Adrshta* is admitted, its significance is not under estimated. Absence of *Adrishta* will result in no contact between the body and self and hence result in collapse of cognition. *Adrshta* seems to be rejected by Nyaya. Kanada makes extensive use of this notion to explain magnetic attraction, initial motion of atoms, falling downwards [10].

## 3.2. 2 *a Priori* Inference

Primarily in the early periods *a priori* inference was preferred for empirical observations. The *a priori* inference can be described as:

Thinking is a property.
A property can reside in a substance alone.
Therefore thinking must be attributed to a substance.
There is no other known substance with thinking as its property.
By elimination, a new substance with thinking as its property must exist [10, page 56].

## 3.2. 3 Observational Inference

Prashastapada who focused on empirical inference redefines inference as observational. It is at two levels as *Drishta* - the observed, and *Samanyato drishta* – the generally observed. Observational inference is illustrated as follows:

Dewlap exists only in cow.
Dewlap is observed in an animal.
Observed dewlap is associated with memory information.
Inference about the observed object is drawn.
Therefore the animal is recognized as cow alone [10, page 66].

Here Prashastapada mentions that the Self has to contact the mind before drawing the final conclusion. A clear parallel can be seen between his method of comparing the unknown object and partly observed object in question with a recollection, to his idea of Self contacting the mind in the final step of inference. Self is the completely known and observed in the physical sense which should contact the mind/*manas* whose existence is known through memory of previous interactions with the same mind. The external process of observation is mirrored in the internal process of understanding. An entire section is dedicated for the establishment of the concept of self expressed as "I".

An illustration for *Samanyato drishta* – The Generally Observed:

As the name suggests this includes substances generally observed like air, which cannot be seen but is inferred through its commonly known properties.

Air possesses touch.
Touch has the attributes of motion and quality.
This substance does not inhere in any other known substance.
Therefore this is a new Substance [10, page 56].

A diagrammatic view of cognition is as follows:

Cognition
↓
Perception ←——————————————→ Inference
↓
*Drshta* ←——————————→ *Adrshta*
↓
↓ Observed ↓ Generally Observed

## 3.2. 4 Empirical Inference

The principle of empirical inference is explained as a cause-effect relation or any of its derivatives. The process of inference is said to occur in one of the following ways:

An object exists ≡ The object of inference exists.
An object exists ≡ The object of inference does not exist.
An object does not exist ≡ The object of inference does not exist [7, page 289].

The inference about a substance can be drawn from both the existence and non-existence of the premise.

The Vaisheshika School does not recognize *Upamana*-analogy and *Shabda*-verbal testimony as ways of acquiring knowledge like the Nyaya School.

# 4 Atoms – The *Anu*

## 4.1 Nature of Atoms

The Vaisheshika sutra about atoms states

*That which is existent and has no cause (i.e., an atom) is eternal. It is not perceived but is inferred from its effect.* [10]. (iv.1.1-5)

Atoms are the primordial infinitesimal particles of everything except space or *Akasha*. To a certain extent terms like atom, space, tend to give us the picture of current-day atom or space, but there are some differences.

Atoms in Vaisheshika are essentially of four kinds: Earth, *Apa*- water, *Tejas*- Fire and *Vayu*-air. These atoms are characterized by their characteristic mass, basic molecular structure such as dyad, triad, etc, fluidity (or it's opposite), viscosity (or its opposite), velocity (or quantity of impressed motion- Vega) and other characteristic potential color, taste, smell or touch not produced by chemical operation. It is these four kinds of atoms involved in all chemical reactions while the space remains unaffected.

## 4.2 Atomic Combinations

Atoms may conjoin or disjoin in reactions. Conjunction and disjunction: Kanada says there are three kinds of conjunction:

a) Contact produced due to motion of one object and not the other.
b) Both may be in motion.
c) Contact by actual contact.

Prashastapada explains the last by referring to an example – consider a dyad of earth which is in contact with two water atoms which are themselves in contact and form a water dyad. Then the earth dyad's contact with the water dyad is produced by the earth dyad's contacts with the water atoms. It is important to note that while one ubiquitous substance like *akasha* (say) may contact non-ubiquitous substance, two ubiquitous substances cannot be in contact since neither is capable of motion. Disjunction is considered by the older Vaisheshikas to be a quality which inheres in a pair of substances when one has just parted contact with the other [10, page 121-122].

## 4.3 Nature of Atomic Combinations

Atoms are invisible though the final substance formed by conjunction of many such atoms is visible. Several causes lead to such a multi-conjunction substance. The atoms unite in pairs and the unification continues until the visible substance is formed. As long as there is no external agent such as heat applied the properties of the atom are reflected in the binary structure as well. The atoms combine driven by an inherent tendency which is their natural property to form dyads. Although Prashastapada seems to have popularized this view of dyads, Kanada's system maintained a different stand.

Kanada says the atoms conjoin as a result of their inherent tendency, but different atoms combine in different patterns. Some in pairs, others in triads, tetrads and so on, which may happen in two different ways,

Atoms combine ≡ basic unit /molecule with two, three or n number of units and not two three or n number of dyads where $n \geq 2$

Further
Basic unit of n atoms ≡ 1 atom + 1 atom… n atoms where $n \geq 1$.

This essentially means

A group of n atoms fall together to form one unit. ($n \geq 1$)

Prashastapada insists on

Atoms combine ≡ only to a binary molecule, not triad, tetrad, etc.

Further
Basic unit of n dyads ≡ 1 dyad + 1 dyad… n dyads where $n \geq 1$

These dyad combinations further combine in different proportions to form isomeric substances. The inherent properties exhibited by these different substances is a result of the collocation process where it may mean quantitative difference or even spatial arrangement since it is only '*paramanu*' generally translated as 'atom' which is a point energy with zero mass and dimension. Therefore the dyad will have a finite mass and size and hence the spatial arrangement too becomes an important qualifier of the properties of the final substance to be formed [9].

This is comparable to the current physics point- of- view of basic particles like electrons, protons, bosons, etc that are mere energy clouds which inter-combine in different combinations to form all the known matter. The properties of energy when treated as a basic unit is a constant, but it is the difference in mere combination in quantity, quality and time that qualifies the finally formed substance as 'The Substance' with its inherent properties. Once the basic unit is formed, further conjunction results from several causes other than the basic impulse or nature of the atom. Hence the atom is different from the substance.

Prashastapada and Kanada concur on the idea of the above said process of combination of the basic unit resulting in the variety of substances which is bound by the laws of universe. Therefore an element of consciousness is considered as playing a role in what the world appears to be.

## 4.4 Atomic reactions

A substance may change qualitatively under the influence of heat in its course of existence. The Vaisheshika's stand on such a change is

Substance A ——application of heat——→ Decomposition in to *paramanus* or the basic unit with zero mass and not the unit of dyad, triad, etc.

↓

Recombination of *paramanus* with a new ←——application of heat basic unit arrangement and order resulting
in a new substance.

The Vaisheshikas hold that under the influence of heat, substances are broken down to the most basic entity (*paramanu*) before being transformed in to a new substance where as the Nyaya school does not believe in decomposition in to the very basic entity. The Vaisheshikas believe that in transformation of a substance the basic properties of the atoms change and the Naiyayikas disagree [9, page 104-105].This also establishes how meticulous the ancient schools and their philosophers were to the very last detail.

Prashastapada gives a specific example for such a reaction. He considers the fertilized ovum under the application of the animal heat or the bio-motor energy.

                    _____germ                    Both are isomeric
                    ↑                              modes of
Fertilized ovum ——action of heat——→                  Earth
                    ↓____ sperm substance

The fertilized ovum breaks down in to its constituents which in turn are reduced in to homogenous earth atoms. They are homogenous because they essentially belong to the same *bhuta*. These basic atoms of the *bhuta* earth re-combine under the influence of the metabolic heat to form the germ-plasm. The germ-plasm develops enriching itself through the nutrients of the body [9].

Germ-plasm —action of heat——→ germ radicals —— action —
                                              ↓
                                   Of         → cells and tissues
                                              ↑
Food substance —action of heat——→ food constituent radicals —— heat —

As can be seen at each stage heat breaks down germ-plasm in to constituent atoms which combine with the constituent atoms of food and all these basic atoms will re-combine to form the cells and tissues. All along heat is a necessary element.

## 5. Hetero-*Bhautic* compounds in action:

The Naiyayikas and Vaisheshikas agree on how atoms of different *bhutas*, i.e. atoms of earth, water, air and fire interact when in contact under the stress of heat. Heat alone is seen as insufficient in many such hetero-*bhautic* reactions and hence a medium is required to keep the reaction going. Hetero-*Bhautic* as the term suggests refers to reactions between atoms of different *bhutas*. A medium is the energizer for the atoms of different *bhutas* in setting up intra-atomic dynamic forces which finally results in a new substance as the end-product of the reaction. It may be easier to understand this with an analogy:

Tea leaves and sugar cannot be chewed together for the effect of tea. Instead, the two have to be boiled in water where water is the substratum, such boiled decoction with or without milk is consumed as tea. Theoretically this is the same as the tea leaves and sugar chewed together.

Likewise certain hetero-*bhautic* atoms require a substratum to inter-bond them though in the end such a substratum remains unaltered. Milk is an example of a quasi-compound where water is the energizer for the earth particles and when water is extracted from milk, milk retains its milky substance [9].

Although all the four *bhautic* atoms can act as the substratum, it is only the earth atoms which can correspond to basic changes in the atoms since, they can arrest the molecular motion which may even be the motion such as liquid flowing due to gravity and the earth atoms can also counteract the tendency of atoms to fall in to a peculiar group or order [9].

## 6. Action of light as a source of heat

### 6. 1 Nature of Light

Vatsayana of the Nyaya School describes the internal heat of a substance as acting in reactions where no external heat is traceable. Such an internal heat is compounded by the solar heat which is the source of all heat stored by substances utilized in chemical changes. An example is given as the change in grass color is due to the *bhuta Tejas*-fire in the form of latent heat which is stored in the atoms of grass obtained from sunlight and not *Tejas* in the form of *Agni* the fire [9]. Such heat cannot be taken away from the substance even by freezing it. It is the same internal heat obtained from the sunlight which ripens a mango fruit eventually resulting in the change of its color, smell, taste not to say about nutritional value.

Light itself is described as constituting indefinitely small particles which mean something smaller and different than so far explained *anu* or *paramanu* since these particles are called neither *anu* nor *paramanu*. These particles radiate themselves in all directions rectilinearly with a conical dispersion from their source and with inconceivable velocity [9, page 105].

### 6. 2 Action of Light

Light can penetrate the inter-atomic or inter-molecular spaces, hence light particles must be indefinitely smaller than the atoms. Such penetration does not always affect the atomic structure necessarily which is explained with an analogy of frying paddy in a pan where the heat from the fire neither affects the pan nor the paddy structure. Such a penetration and passing through of the light rays accounts for translucent and transparent objects. In other instances the light particles might rebound off the atomic surface as in reflection or simply are obstructed like in opaque objects to cast the shadows of the objects [9].

All these processes also indicate no decomposition or re-composition of the atomic structure of the object of incidence.

## 7 *Parispanda* – The fundamental Motion

The atoms possess an inherent rotary or vibratory motion – *Parispanda* which is the root cause of all visible or invisible action and operation involving matter. Only an ideal body is devoid of any motion [6]. *Akasha* –The Space is devoid of this vibratory motion since it has no atomic structure. *Vayu* which is matter in gaseous form is explained as a state of *Parispanda* in action. The Nyaya-Vaisheshika holds such *Parispanda*-the basic inherent motion of all atoms as the basic form of all activity in the universe because all existing matter can be reduced to such jiggling atoms. The philosophers of this school also add that the concept of "Maheshvara" or any such godly agent is included to satisfy the philosophers but not to undermine either the power of *Parispanda* or the material causes and effects as can be seen or perceived or inferred. Yet consciousness is excluded from those substances which are affected by physical motion of atoms. At this juncture it is important to note that although the invisible god is not seen as the cause of all unexplainable phenomena the invisible consciousness is not treated similarly. The importance of consciousness is as real as any real visible object.

A problem: A fundamental problem for Vaisheshikas was to explain how imperceptible atoms could combine to produce perceptible individuals. There is textual evidence to suggest that the early Vaisheshikas held the straight forward view that surely if one takes enough things below the threshold of perception and sets them beside each other it will produce something perceptible [10].

## 8. Matter Classification

### 8.1 Introduction

Matter classification begins with classifying everything in to *Dravya*, *Guna* and *Karma* roughly translated in to Substance, Quality and Action. Such a translation is limited by language since even technical terms are influenced by the philosophy of the culture they originate in. A categorization such as this indicates an organized discipline of study. A clear differentiation is made between matter and its attributes. Here *Dravya* – the substance is a term used in a narrow technical sense to exclude quasi-compounds which refers to compounds made of atoms from the different *bhutas*, i.e. poly-*bhautic* compounds.

### 8.2 *Dravya* – The Substance

The *dravya* which is all that can be perceived is sub-classified in to five physical substances as earth, air, water, fire and space which are the *Pancha Bhutas*.

Substance is defined as the material cause of its quality and action. Kanada deals with each of the sub-categories of substance through their properties like: color, taste, smell associated with earth; color, taste, fluidity with water; so on with air and fire. It can be understood as:

| Properties | Color | Taste | Smell | Touch | Fluidity | Viscosity |
|---|---|---|---|---|---|---|
| 1. Earth | * | * | * | * | — | — |
| 2. Water | * | * | — | * | * | * |
| 3. Fire | * | — | — | * | — | — |
| 4. Air | — | — | — | * | — | — |

Table 1

Fluidity is observed to occur both naturally and instrumentally. Water is in a liquid state naturally and can be converted in to gaseous state or solid state by externally applying or removing heat. Yet fluidity will be the primary state of water. Likewise butter or ghee is normally in a solid state (at least in winter) but can be melted in to a liquid state with very little application of heat. In such a case solid state would be their primary state [1].

*Dravya*, *Guna* and *Karma* give rise to further categories. All the three are non-eternal. They are explained through their properties. *Dravya* possesses in it both *guna* which is quality and *karma* which is action, e.g. Fire which is a *dravya* possesses both the *guna* of heat and the action of moving upwards as in a flame. Thus *Dravya*, *Guna* and *Karma* have a real objective existence since they are associated with the real substances.

## 8.3 *Karma* – The Motion

*Karma*, motion, is a deeper concept than mere physical displacement with respect to time. Kanada defines five kinds of motions. They are *Utksepana*- ejection, *Avaksepana*- attraction, *Akunchana*-contraction, *Prasarana*- expansion and *Gamana*- composite movement. Vyomashiva clearly explains that motion is not instantaneous instead it is incremental. This is true even in a process like cooking the food where the food is neither cooked instantaneously nor does a change occur in its state until a minimum energy is expended [1, page 22]. Such a minimum energy can be seen as similar to the threshold energy concept of today. Here the threshold energy is greater than the rest energy of the final product to be obtained instead of a particle in the concept of current physics. The incremental nature of change in substances explained by Vyomashiva is what follows from today's relativistic physics about no action being instantaneous.

## 8.4 *Akasha* – The Space

Space or *Akasha* is defined as that which has none of these attributes. Space is recognized as a separate category and it is clearly not earth's atmosphere. It is conspicuously eliminated from the sub-categories of *Dravya* which constitute atoms and it is an entity like a being is an entity. *Akasha* or space has no atomic structure and it is consequently inert.

In very early times, ancient Indians had visualized the spherical nature of space and concluded it is the earth's rotation which causes day and night [1].

## 8.5 Space - Time

It is interesting to note the background in which the ancient Indians propose the existence of Space and Time as entities. Bhaduri one of the philosophers summarizes this insightfully as – "We perceive pairs of objects with qualities of remoteness and nearness, spatial or temporal – inhering in them. Furthermore, we are able to make comparative judgments of this sort – we can say A is farther from B than from C, etc. What enables us to make this judgment? It is the greater number of contacts between individuals spread out between A and B than between A and C. For example, the ink bottle is nearer to the pen than to the radiator, that is, the number of contacts present in a line from the ink bottle to the radiator is greater than the number from the ink bottle to the pen. Only thus can the notion of "greater distance" be explained. But when we look for individuals whose contacts must be counted up, we do not find any belonging to other categories – or at least we do not find the right number to describe such nearness or farness. Between the inkbottle and the pen a book (say) is situated, while between the inkbottle and the radiator there is just space! Thus, in order to provide the material to explain these comparative judgments we must postulate an intervening series of entities and these are spatial." As Bhaduri puts it, contact is not a transitive relation instead space is introduced to make it transitive and more generally to relate two otherwise unconnected things by a series of contacts postulated to lie between. This is spatial discrimination [10].

Likewise Vachaspati states that for the concept of A is older than B the entity of time needs to be introduced. Therefore there is a particular spatial temporal relation connecting each pair of objects. Then why aren't there as many spaces and times as there are relations of this sort? Vaisheshika talks of one space and one time since all objects in this continual space and time can be inter-related. If there were more than one space and one time, there is a possibility that an object from a certain space/time cannot be linked with another object from a different space/time [10].

The Vaisheshika atomic substances are defined in a matrix of four non-atomic substances: Time, Space, Soul and Mind [1].

A map of sizes, eternality and nature can be as follows:

| Substance | Size | Eternality | Nature |
|---|---|---|---|
| Atom | atomic | Eternal | Active |
| *Manas* | atomic | Non-Eternal | Active |
| Time | *mahan* | Eternal | Active |
| *Akasha* | *mahan* | Eternal | Inactive |
| Soul | *mahan* | Eternal | Active |

Table 2

The size atomic and *mahan* correspond to spherical structure being small and big respectively. Chandramati states that the same spherical nature resides in an atom when in minute proportion and in *Akasha*, time, place and self in infinite proportions. Space as a *dravya* that is *Akasha* has no absolute properties since space and time are relative in Indian science [1, page 26].

## 8. *Samanya, Vishesha* and *Samavaya*.

The next three categories in the total six categories of Vaisheshika are *Samanya, Vishesha* and *Samavaya. Samanya* is the class concept and *Vishesha* is particularity. Kanada's view about *samanya* and *vishesha* is different from that of the later philosophers.

Let us begin with Kanada. It has been suggested that translating these two terms as "genus" and "species" would render Kanada's intent most accurately. Kanada makes a statement that a *samanya* may also be viewed as a *vishesha* in cases other than the Being. This can be understood by applying this concept to an entity such as, e.g., potness, which is a *samanya*-genus relative to particular pots but a species relative to the more inclusive genus clay-objectness.

The later thinker's concept of universals is that they are real, independent, timeless, ubiquitous entities which inhere in individual substances, qualities and motions and are repeatable, i.e., may inhere in several distinct individuals at once or at different times and places. The general term used for such an entity is *samanya*. Either way the postulation of universals in Vaisheshika calls for the necessity of explaining the existence of natural kinds, the fact that certain entities are similar because of a true similarity and not merely because we think so [10, page 133-134].

A special feature of this school is *Samavaya* which is inherence referring to the special feature of individual atom of even the same element. Kanada explains inherence as the cause of notion that something is "here" in a locus and connects its function to causality. He also conceives that there is only one inherence, since there is no indication that different inherences connect different pairs of things related by inherence [7, page 292].

*Samanya* , *Samavaya* and *Vishesha* are products of intellectual discrimination [1]. Such an idea is further emphasized by Kanada by including only *Dravya, Guna* and *Karma* under *Bhava* the Being.

Time is defined as the cause of all non-eternal things and time is irrelevant and therefore absent for eternal substances. In Vaisheshika universal is taken to be ubiquitous and timeless. Whatever can be defined with respect to time and space cannot be a universal. The process that marks the passage of time on an object will thus be relative.It is only the universal which is true for all time and space and it is the being [1]. Time is clearly differentiated from space and space is not absent from eternal things. This time is associated with motion, which begins with the universe in its cyclic life of creation and dissolution. The time is at rest when the universe is in the process of creation. An analogy of this is working with a piece of clay trying out various combinations and figures in it and making such arrangements with it as will lead to certain ends and aims which are potentially in it inherently.

## 9. The Being or Existence

Kanada and the early Vaisheshikas view existence as the highest genus which is not a genus species lying under any superior genus. Kanada calls such a supreme universal as '*Bhava*' from the root *bhu* meaning "to come to be", and he specifically mentions that *bhava* includes *dravya*, *guna* and *karma*. Such a *bhava* is called *Satta* by the time of Chandramati and Prashastapada [10]. This indicates a clear differentiation between material or real concepts and intellectual or transcendental ideas even at the time of Kanada. The inclusion of consciousness or the Being as a variable in the studies of all matter or materialistic sciences does not hinder the understanding of matter, instead adds a new dimension to the perspective.

## 10. Conclusion

The Nyaya-Vaisheshika begins with the beginning of universe in an endless cycle of existence. This existence is marked by all sorts of motions- motion as microscopic as the inter-atomic vibrations and motion as macroscopic as that of the planets and stars. The motion ceases only during the rest period when the universe is preparing for the next cycle and at such a time all the atomic laws collapse. The universe can be understood as guided by the will of a personified god or the laws of nature *Rta* at an abstract level. The knowledge of universe, its creation, dissolution and life is comprehended by the consciousness which is an active element in all the actions of universe.